# 3D cosmic-ray muon tomography using portable muography detector

Kullapha Chaiwongkhot[1,2], Tadahiro Kin[1], Yuta Nagata[1], Tomohiro Komori[1], Naoya Okamoto[1], and Hamid Basiri[1]

[1] Interdisciplinary Graduate School of Engineering Sciences, Kyushu University, Fukuoka, Japan
[2] Department of Physics, Mahidol University, Bangkok, Thailand

E-mail: kullapha_ch@kyudai.jp

**Abstract**

A feasibility demonstration of three-dimensional (3D) muon tomography was performed for infrastructure equivalent targets using the proposed portable muography detector. For the target, we used two sets of lead blocks placed at different heights. The detector consists of two muon position-sensitive detectors, made of plastic scintillating fibers (PSFs) and multi-pixel photon counters (MPPCs) with an angular resolution of 8 msr. The maximum likelihood-expectation maximization (ML-EM) method was used for the 3D imaging reconstruction of the muography simulation and measurement. For both simulation and experiment, the reconstructed positions of the blocks produce consistent results with prior knowledge of the blocks' arrangement. This result demonstrates the potential of the 3D tomographic imaging of infrastructure by using eleven detection positions for portable muography detectors to image infrastructure scale targets.

Keywords: Muon tomography, Cosmic-ray muon radiography, Muography, Cosmic rays, Muon detector

## 1. Introduction

Cosmic-ray muon radiography, also known as muography is one of the effective techniques for nondestructive radiography. The technique has been used to survey the structure of huge objects with the basic principle similar to conventional X-ray imaging. The muography has been used in various fields, including the finding of a void in the Khufu's pyramid [1], prediction of volcano eruption [2], and investigation of the melted core of the Fukushima Dai-Ichi Nuclear Power Plant [3]. However, only a few applications have published the three-dimensional (3D) muography based on the attenuation method. Besides, all of these results are focused on geological objects, such as Asama Volcano[4, 5], Showa-Shinzan Lava Dome [6], ore exploration[7], and geological mapping [8, 9].

The application has expanded to smaller-scaled objects such as degradation surveys of social infrastructures, including concrete-based bridges, furnaces, dams, and valuable architecture. Portable muography detectors [10, 11] for smaller-scaled targets have been developed from their operation and maintenance. The muography has the potential to estimate the density variance caused by degradation in the object obtained two-dimensional (2D) projection, showing only the variance in density integrated along the transmission path of muons, i.e., the spatial distribution along the path cannot be estimated. Thus, the tomography technique, such as filter back projection, ordered subsets expectation maximization (OSEM), and maximum likelihood-expectation maximization (ML-EM) method used in the medical field, is necessary to yield the spatial distribution inside the objects. However, direct measurement of the overall 3D image is impractical due to the low flux of cosmic-ray muons, which is approximately 1 muon cm$^{-2}$ min$^{-1}$ at sea level.

Moreover, the flux is drastically decreased in proportional to the cosine square. Particularly, when an object is surveyed in 3D, the measurement time and location are restricted. A novel image reconstruction algorithm is strongly required to solve the ill-posed problem.

In the present work, we propose an approach based on the ML-EM algorithm [12] for cosmic-ray muon tomography. A simulation was developed to test the feasibility of the algorithm [13] and designed the experiment's configuration using our developed portable muography detector [10]. Then, the 3D muography image from the experiment setup was also performed to confirm the feasibility.

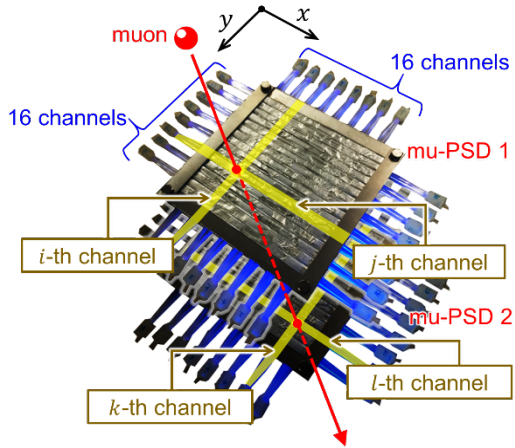

**Figure 1.** Photo of two muon position-sensitive detectors fabricated in the portable muography detector. The detected positions of a muon trajectory are recorded as MPPC ID numbers (i,j) and (k,l) for upper and lower mu-PSDs.

## 2. Materials and methods

### 2.1 Portable muography detector

A multi-purpose portable muography detector was developed for real-time monitoring of the degradation of infrastructures. We successfully mapped the inner structure of a seven-story concrete building using a detector with an adjusted 8 msr angular resolution [10]. The portable muography detector was also used in this measurement. Figure 1 depicts a schematic view of the detector. Each muon position-sensitive detector (mu-PSDs) is a combination of two layers of PSFs: Kuraray, SCSF-78, 2.0 mm SQ, BJ) arrays. The layers are placed perpendicularly to each other to detect hitting position of muons. The scintillation light generated by the incident cosmic-ray muons is propagated along the PSFs and detected by sixteen MPPCs connected on one side of the layer. The muon hitting positions are recorded as the MPPC ID numbers $(i,j)$ and $(k,l)$ for the upper and lower mu-PSDs, respectively.

The mu-PSDs are installed in a heat-insulating box where the temperature inside was controlled to 17 °C using a Peltier heating & cooling device (Ohm Electric Inc., BOXCOOL, OCE-F40F-D24) to keep the gain shift of the MPPCs at a low level. The detector properties such as stability, detection efficiency, and several measurement results are described in Ref. [10].

The projection of an object is evaluated from the cosmic-ray muon's absorption ratio distribution as a function of muon direction in muography based on absorption method. The direction is defined by the zenith angle and the azimuth angle; $(\theta, \varphi)$. This detector generally requires background and foreground measurements to derive the absorption ratio necessary for general muography detectors utilizing the absorption method. Given a background intensity $I_0(\theta, \varphi)$, the absorption ratio $R(\theta, \varphi)$ for measured intensity $I_t(\theta, \varphi)$ can be expressed as follows:

$$R(\theta, \varphi) = 1 - \frac{I_t(\theta, \varphi)}{I_0(\theta, \varphi)}, \quad (1)$$

The output data of the detector is discretized as $R(\Delta x, \Delta y)$, where $\Delta x$ and $\Delta y$ represent integers calculated by $\Delta x = k - i$ and, $\Delta y = l - j$ for a muon hitting positions $(i,j)$ and $(k,l)$, respectively. In the present paper, the authors refer to the absorption ratio distribution as a vector plot.

### 2.2 Three-dimensional cosmic-ray muon tomography using ML-EM algorithm

The ideal detector for 3D tomography has a solid angle of $4\pi$ sr to the target. As Radon's theorem [13] proved, the target shape can be determined uniquely under this condition. However, as mentioned in the Introduction section, the effects of low cosmic-ray muon flux and its angular distribution on the measurement duration and the number of detection positions are significant constraints in muography. The constrain contributes to the demand for optimization between the shortest possible measurement duration, a reasonable statistical error of measurement, and the capability to reconstruct the 3D image. The numerical methods are required to solve this ill-posed problem. Like the ML-EM method, the iterative algorithm is well-known for its successful implementation in "emission tomography" such as SPECT or PET scans to reduce measurement time, which has similar constraints to the muography. Thus, we adopted this method to improve the 3D cosmic-ray muon tomography. The ML-EM algorithm estimates the 3D density $\lambda(b)$ of a voxel $b$ that maximizes the likelihood of the projection $n'(d)$ of pixel pair $d = (i,j), (k,l)$. In this study, $n'(d)$ was intended to be equivalent to the absorption ratio $R(\Delta x, \Delta y)$ of the vector plot.

According to the flow chart in Figure 2, the algorithm can be described in the following steps:

(i) The ML-EM method begins with an initial estimation of the $\lambda^{(0)}(b)$ of a voxel $b$ where $b= 1, 2,...,B$ and $B$ is the number of voxels in the region of interest. The superscript represents the number of iteration loops which is zeroth at this initial step.

(ii) The projection $n'(d)$ is calculated from the initial estimation by

$$n'(d) = \sum_{b=1}^{B} \lambda^{(0)} p(b,d) \quad (2)$$

where $p(b,d)$ refers to the probability of detecting a muon event from the voxel $b$ by the pair-pixels $d$ where $d = 1,2,...,D$. Here $D$ denotes the total number of muon directions that can be detected by the detector pair-pixels. In this study, $p(b,d)$ is estimated by the distance-driven back projection method [14]. The method calculates the probability from the projected area ratio of the voxels to the detector pair-pixels onto a common plane.

(iii) Assuming that $\lambda^{(m)}(b)$ denotes the estimated density in a voxel $b$ at the $(m)^{th}$ iteration. The next approximation $\lambda^{(m+1)}(b)$ can be expressed by the ML-EM algorithm as proposed by Shepp and Vardi [15];

$$\lambda^{(m+1)}(b) = \lambda^{(m)}(b) \left\{ \frac{1}{\sum_d p(b,d)} \sum_d \frac{n(d)p(b,d)}{n'(d)} \right\}. \quad (2)$$

(iv) The updated projection $n'(d)$ of the $(m+1)^{th}$ iteration can simply be derived by the convolution process in equation (2).

(v) The updated $n'(d)$ was compared with the experimental absorption ratio $n(d)$, which is the absorption ratio $R(\Delta x, \Delta y)$ for this muography method. This step is the criteria to continue or stop the iteration.

In Figure 2, the $(M)^{th}$ iteration indicates the last iteration that provides the finalized 3D density image $\lambda^{(M)}(b)$. However, the iteration stopping criterion is required for the acceptable quality. In this study, a simulation code was developed to estimate the number of iterations (See [13] for detail).

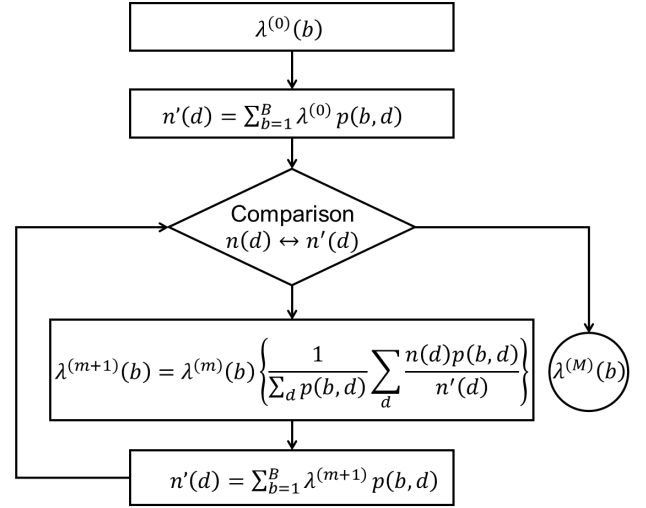

**Figure 2.** A flow chart of the ML-EM algorithm.

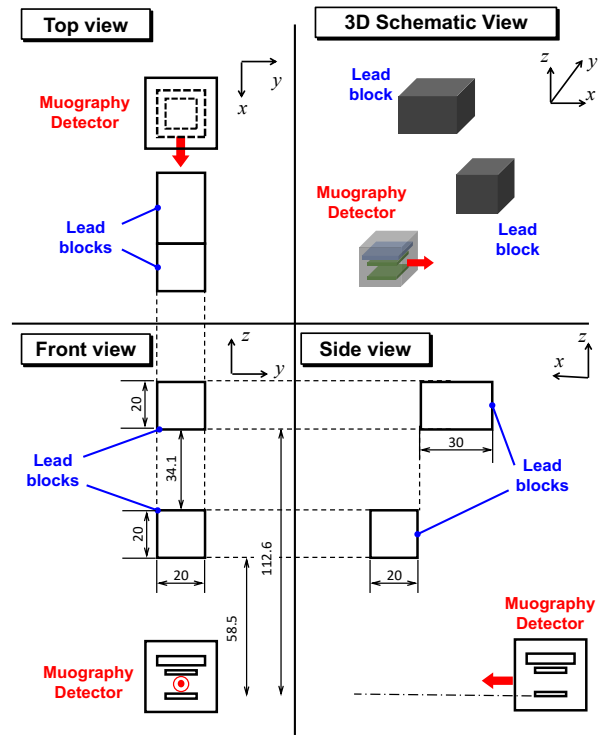

**Figure 3.** The sketch of the set of the lead blocks configuration for the experiment.

## 3. Experiment

### 3.1 Measured objects

For the feasibility study, first, we used lead blocks to reproduce the absorption ratio of the infrastructure buildings because the whole system can be reduced 4.4 times based on the density difference. Particularly, 20 cm of lead thickness corresponds to a five- or six-story building with a 15 cm thick floor of concrete. Therefore, we placed two different sized rectangular lead blocks ($20 \times 20 \times 20$ cm$^3$ and $20 \times 20 \times 30$ cm$^3$) at extra height (58.5 cm for smaller and 112.6 cm for larger lead blocks), as shown in Figure 3.

*3.2 Simulations to determine experimental conditions*

The simulation study determined the detection positions, measurement times, and the ML-EM convergence conditions (See Ref. [13] for the detail). Figure 4 depicts the considered coordinations of detector positions. The origin was set at the center of the mu-PSDs of the 3$^{rd}$ position and the center of the lower lead block. The absorption ratio maps were calculated for all eleven positions using a Monte Carlo method. The number of events was 10$^7$ muons, with a statistical uncertainty of less than 1% for all pixel pairs.

We performed the ML-EM analysis using the simulated data by changing the number of detection points and the iteration number in the reconstruction procedure to find optimal ones. Figure 5. Simulation results for optimizing the number of detections point and iteration number in the ML-EM method. Dashed and solid lines show the lead blocks region. shows the detector position, located along the x-axis for seven different positions, and the others are varied along the *y*-axis. The coordinate (*x*[cm], *y*[cm]) represents the position of the center of the mu-PSD2 in each simulation. The vector plot was simulated for each of these 11 detection positions to minimize the measurement time. Then, the optimized number of detection positions for the practical measurement was designed from the quality of the reconstructed 3D image.

The ML-EM method was applied to the image reconstruction of the simulation data. Various combinations of the number of the detection positions and the iteration "*k*" were investigated to optimize the experimental condition, v. Figure 5 shows the 3D images projected to the *xz* plane in different conditions. The number of the detection position of 3, 5, 7, and 11 in the figure consists of the detection positions 2–4, 1–5, 0–6, and 0–10, respectively. The detection position number, indicated by the bold number in the image of the two lead blocks, became apparent with the increase in the number of detection positions and iteration. Without the projection from the various detection position along the *y-axis*, the position of the upper block is not clear. However, for the seven detection positions, the result is reasonable for finding the center position of two lead blocks. Thus, seven detection positions were chosen with 20 iterations for the 3D image reconstruction of the measurement result.

In this study, only the size and location of the lead blocks were , i.e., the muon interaction in the matter was out of the simulation scope. Furthermore, as the attenuation rate is proportional to the object's thickness, the vector *n(d)* in the simulated image reconstruction was determined by the simulated average thickness instead of the attenuation rate.

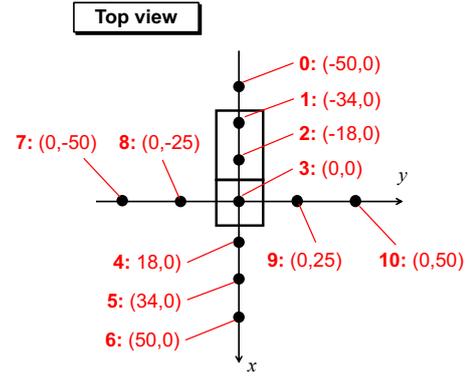

**Figure 4.** Detection positions for 3D cosmic-ray muon tomography in a unit of cm. The coordination represents the center of mu-PSDs given the origin at the 3$^{rd}$ measurement position.

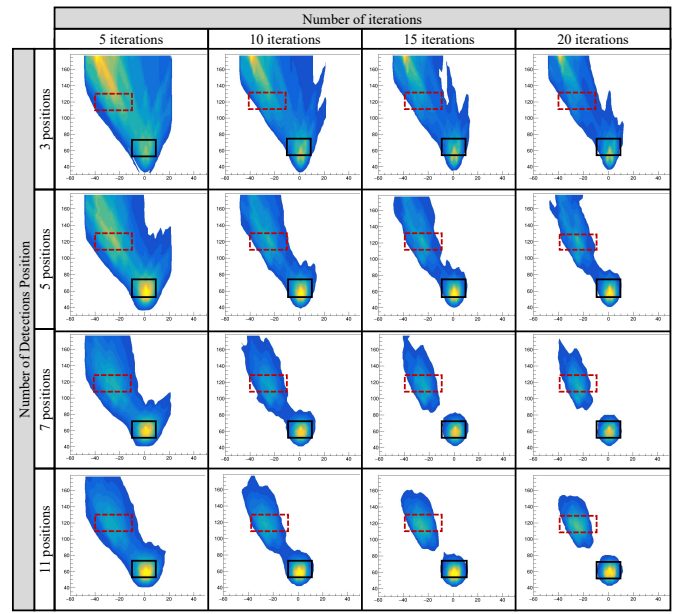

**Figure 5.** Simulation results for optimizing the number of detections point and iteration number in the ML-EM method. Dashed and solid lines show the lead blocks region.

**4. Data analysis**

First, Eq. 1 was used to calculate absorption ratio maps for each detection point, $R_0(\Delta x, \Delta y)$ - $R_6(\Delta x, \Delta y)$, using measured intensities, $I_{t0}(\Delta x, \Delta y)$ - $I_{t6}(\Delta x, \Delta y)$ and background intensity $I_0(\Delta x, \Delta y)$ that were previously measured for a common one. Figure 6 (a) is an example of the thickness project result. The simulation provided the two lead blocks thickness projection without the artifact noise.

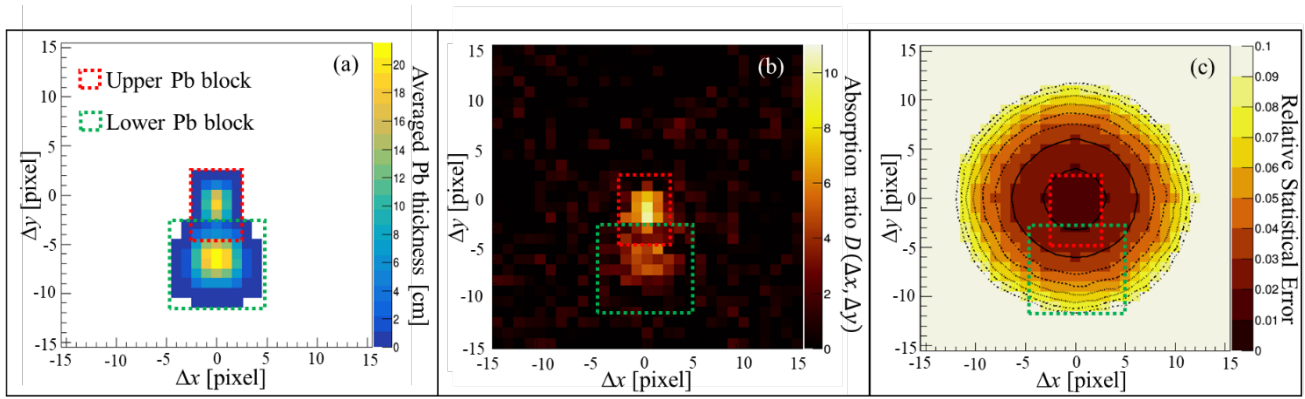

**Figure 6.** (a) Lead thickness along to directions of the vectors from pixel in bottom mu-PSD to that in top mu-PSD. Absorption ratio of measured data (b) is consistent with (a). Experimental statistical uncertainty is shown in (c) and is kept lower than 10% where the lead block region.

The simulation could identify the projection pixels that contain the information of two lead blocks. Figure 6 (b) depicts the attenuation rate mapping derived from the same setup measurement. In contrast, the measured result contained the noise, which corresponds to the relative statistical uncertainty distribution in figure 6 (c). Thus, the simulated projection was used to distinguish the blocks' data contained pixels from the noise.

The duration measurement for each position was evaluated using the simulated projection and relative statistical uncertainty. Each position measurement was designed to stop when the whole block projection had less than 7% relative statistical uncertainty. For example, in figure 6 (c), the measurement at position ID 2 was halted at this relative uncertainty level. Table 1 presents the duration of measurement for each position. Only seven detection positions were performed, as shown in Table 1. The determination of the number of practical detection positions is described in the next section. Moreover, the cosmic-ray muon background measurement was carried out for 208 h without the lead blocks.

**Table 1.** The detection positions and durations

| Position ID | Position (x, y, z) [cm] | Duration [hours] |
|---|---|---|
| 0 | (−50, 0, 0) | 407 |
| 1 | (−34, 0, 0) | 308 |
| 2 | (−18, 0, 0) | 166 |
| 3 | (0, 0, 0) | 89 |
| 4 | (18, 0, 0) | 164 |
| 5 | (34, 0, 0) | 344 |
| 6 | (50, 0, 0) | 334 |
| 7 | (0, −50, 0) | - |
| 8 | (0, −25, 0) | - |
| 9 | (0, 25, 0) | - |
| 10 | (0, 50, 0) | - |

Due to the higher statistical uncertainty from the experiment than the simulation, the relative statistical uncertainty was used as a threshold to filter out some pixels of the measured vector plot in the ML-EM algorithm of the experiment results. If the relative statistical uncertainty of the pixels is greater than 7%, the pixels were ignored by the imaging algorithm. The threshold level was set based on the experimental condition of measurement duration. The uncertainty is also used to weight the probability parameter; $p(b,d)$ to decrease the affection of the higher uncertainty pixel.

## 5. Results and discussion

Figures 7a and 7b present the reconstructed positions of the lead blocks in 3D from the simulation and experiment, respectively. The data of seven detection positions were included in the 20 iterative calculations, and the voxel size was 2 cm by 2 cm. The experimental 3D muography also indicated the visible noise around the blocks' position. However, for both simulation and experiment, the image reproduces reasonably well the position of the upper and lower lead blocks, as shown in the red line and black line, respectively. The center of mass calculated from the image vector $\lambda^{(m)}$ (7b) based on simulation is located at (0, 0, 65) for the upper block and at (−26, 0, 122) for the upper block. The center of mass obtained from experimental data is located at (2, 0, 68) for the upper block and at (−22, 0, 122) for the upper block. The 3D image reconstructed from simulation certified the accuracy of the developed ML-EM algorithm. The overestimated size of the upper blocks along the z-axis is reasonable for detection positions where they are located only on the $xy$ plane. The reason why the centers of mass of both 3D images are displaced from the actual setup at (0, 0, 68) for the upper block and at (−25, 0, 123) for the upper block is due to $xy$ plane constraint. Remarkably, there is no initial bias input in this reconstruction. These reconstructed 3D images are calculated from the uniform initial estimation of the image vector $\lambda^{(0)}(b)$ value of 1 for each voxel.

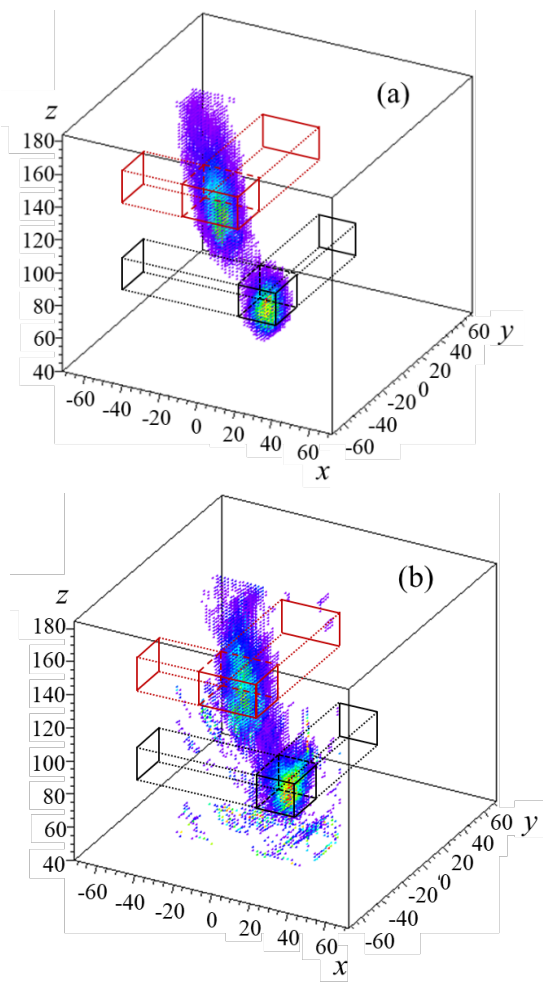

**Figure 7.** 3D muography; (a) from simulation and (b) from experiment.

## 6. Conclusion

In this study, the ML-EM method was used for the 3D image reconstruction of cosmic-ray muon tomography based on the transmission method. The feasibility study was performed under limited detection angles and the acceptance of a portable muography detector. A portable muography detector was used to detect two sets of lead blocks placed at a different heights to the detector. Monte Carlo simulation was used to optimize the measurement duration and confirm the accuracy of the ML-EM algorithm. Finally, the 3D muography image was reconstructed using the ML-EM method from the simulation and the attenuation rate projection obtained by the peer set up measurement. It was found that the image reproduces the position of the two blocks reasonably well. The experimental 3D image confirmed the feasibility of the 3D muography used in our portable muography detector on the infrastructure scale target. The iteration stopping criteria of the image reconstruction will be determined in the near future.